\pdfoutput=1
\documentclass[journal]{IEEEtran}

\usepackage{cite}
\usepackage[pdftex]{graphicx}
\usepackage{amsmath}
\usepackage{amsthm}
\usepackage{amssymb}
\usepackage{algorithm}
\usepackage{algpseudocode}
\usepackage{array}
\usepackage[caption=false,font=footnotesize]{subfig}
\usepackage{amsfonts}
\usepackage{optidef}
\usepackage{bookmark}
\usepackage{float}
\usepackage{tabularx}
\usepackage{slashbox,pict2e}

\newtheorem{theorem}{Theorem}
\newtheorem{lemma}{Lemma}

\begin{document}

\title{Iterative method for simultaneous sparse approximation}
\author{Sahar~Sadrizadeh,
        Shahrzad~Kiani,
        Mahdi Boloursaz,~\IEEEmembership{Student~Members,~IEEE,}
        and~Farokh~Marvasti,~\IEEEmembership{Senior~Member,~IEEE}
\thanks{S. Sadrizadeh, S. Kiani, M. Boloursaz and F. Marvasti are with the Advanced Communication Research
Institute (ACRI), Electrical Engineering Department, Sharif University of Technology,
Tehran, Iran (email: sadrizadeh\_s@ee.sharif.edu; shkianid@gmail.com; marvasti@sharif.edu).}}

\markboth{Journal of \LaTeX\ Class Files,~Vol.~14, No.~8, August~2015}%
{S. Sadrizadeh and S. Kiani : Simultaneous Sparse Approximation}

\maketitle

\begin{abstract}
This paper studies the problem of Simultaneous Sparse Approximation (SSA). This problem arises in many applications which work with multiple signals maintaining some degree of dependency such as radar and sensor networks. In this paper, we introduce a new method towards joint recovery of several independent sparse signals with the same support.  We provide an analytical discussion on the convergence of our method called Simultaneous Iterative Method with Adaptive Thresholding (SIMAT). Additionally, we compare our method with other group-sparse reconstruction techniques, i.e., Simultaneous Orthogonal Matching Pursuit (SOMP), and Block Iterative Method with Adaptive Thresholding (BIMAT) through numerical experiments.  The simulation results demonstrate that SIMAT outperforms these algorithms in terms of the metrics Signal to Noise Ratio (SNR) and Success Rate (SR). Moreover, SIMAT is considerably less complicated than BIMAT, which makes it feasible for practical applications such as implementation in MIMO radar systems.
\end{abstract}

\IEEEpeerreviewmaketitle

\section{Introduction}\label{introduction}
\IEEEPARstart{S}{parse}
 signal processing has recently been exploited in various fields of communication, due to fact that sparse signals can be approximated by only a few nonzero coefficients and hence sub-Nyquist sampling and Compressed Sensing (CS) \cite{candes2006robust,donoho2006compressed}. The general CS problem is formulated as follows:

\begin{mini}|l|
  {}{\|\mathbf{x}\|_{0}}{}{}
  {\label{eq:1}}
  \addConstraint{\mathbf{y}=\mathbf{A}\mathbf{x}+\mathbf{v}}
\end{mini}
where x is the main sparse signal, y is the measurement vector, A is the sensing matrix, and v is the additive noise vector. Two main models are considered in CS for reconstruction of sparse signals. Models with one measurement vector are referred to as Single Measurement Vector (SMV) models, while the other models with at least two measurement vectors are called Multiple Measurement Vector (MMV) models.

    The problem investigated in MMV models, known as SSA, aims to jointly recover sparse representation of the measurement vectors. The SSA applications may be encountered in various fields such as sensor networks \cite{duarte2009model,gastpar2003source}, Electroencephalography and Magnetoencephalography (EEG and MEG) \cite{gorodnitsky1995neuromagnetic}, source localization \cite{malioutov2005sparse}, and distributed MIMO radar systems \cite{ibernon2008comparison}.

    \cite{mohammadi2017sampling} investigates the theory of MMV models. Some algorithms have been developed by extending the general SMV model into the MMV model to solve the SSA problems. Orthogonal Matching Pursuit (OMP)\cite{tropp2007signal} as a greedy algorithm is one of the very first algorithms used for sparse recovery. At each iteration of this algorithm, the best local improvement to the current approximations is found in hope of obtaining a good overall solution. The extension of the OMP algorithm to The MMV paradigm, Simultaneous OMP or SOMP, has been developed in \cite{determe2017noise,tropp2006algorithms,rakotomamonjy2011surveying}.

    The Iterative Method with Adaptive Thresholding (IMAT) algorithm was originally proposed for sparse signal reconstruction from missing samples \cite{marvasti2012sparse,marvasti2012unified,marvasti2012nonuniform}. The Block Iterative Method with Adaptive Thresholding (BIMAT) \cite{abtahi2016iterative} as an extension of IMAT is employed for block sparse recovery for distributed MIMO radar systems.

    In this paper, we propose SIMAT as a generalization of IMAT for simultaneous reconstruction of jointly sparse signals from their missing samples.

\subsection{Paper Overview }
    The rest of this paper is structured as follows. In Section \ref{method}, we first provide the description of SSA model. Then the proposed method is introduced and its convergence is analyzed. Numerical experiments of our method in comparison with the SOMP algorithm are presented in Section \ref{simulation}. SIMAT is then demonstrated as a simple decoding algorithm for MIMO radar systems, and its performance is compared with BIMAT by means of simulation. Finally, the paper is concluded in Section \ref{conclusion}.

\subsection{Notations}
Scalar variables, vectors, and matrices are denoted by italic lower-case, boldface lower-case, and boldface upper-case, respectively. The elements of a vector are denoted by subscript, i.e., $x_{i}$ is the $i$-th element of the vector $\mathbf{x}$. $|\mathbf{x}|$ calculates the absolute value of each entries of the vector $\mathbf{x}$. The pseudoinverse of matrix $\mathbf{A}$ is represented by $\mathbf{A}^\dagger$.  Finally, the output of the thresholding operator $\rm{TH}( \mathbf{x},thr)$ is defined as a diagonal matrix whose diagonal entries are determined as follows:

\begin{equation}\label{eq:2}
{\rm{TH}}{\left( {{\mathbf{x}},thr} \right)_{ii}} = \left\{ {\begin{array}{*{20}{c}}
{1.\;\;\;\;\;\;\;\left| {{x_i}} \right| \ge thr}\\
{0.\;\;\;\;\;\;\;\left| {{x_i}} \right| < thr}
\end{array}} \right.
\end{equation}

\section{The Proposed Method}\label{method}

\subsection{Problem Statement}
    In this section we provide the formulation of SSA problems. Assume that $\mathbf{x}^1, \mathbf{x}^2, ..., \mathbf{x}^L$ are $L$ equal-length signals that share the same sparsity support in a specific transform domain. These signals are random-sampled in another domain by independent sampling masks $\mathbf{s}^{1}, \mathbf{s}^{2}, ..., \mathbf{s}^{L}$. These masks are binary and each element of them is generated independently based on a Bernoulli distribution , i.e., ${\mathbf{s}^{i}}_n\sim Bernoulli(p)$, $ 1\leq i \leq L$, and $ 0\leq p \leq 1$ is the sampling probability. The random-sampled signals $\mathbf{y}^{1}, \mathbf{y}^{2}, ..., \mathbf{y}^{L}$ are derived as follows:

\begin{equation}\label{eq:3}
{y^i}_n = {s^i}_n \times {x^i}_n
\end{equation}

   The problem is to simultaneously reconstruct the original sparse signals from their random-sampled versions by employing the additional information that the signals share a common support in a specific transform domain.

\subsection{Algorithm}
One can find the SIMAT algorithm in Algorithm \ref{alg:1}. Let $n , m$ and $k$ denote the lengths of the original signal, the length of the observed vectors, and the sparsity number, respectively. Moreover, let $\mathbf{x}^{j}_i$ represent the reconstruction of the $j$-th signal after $i$-th iteration.

    It should be noted that the measurement matrices, i.e., $[\mathbf{A}^{1}, \mathbf{A}^{2}, ..., \mathbf{A}^{L}]$, can be calculated by multiplying the transformation matrix, which maps the signals to their sparsity domain, and diagonal matrices whose diagonal entries are equal to the elements of  $\mathbf{s}^{1}, \mathbf{s}^{2}, ..., \mathbf{s}^{L}$.

    This algorithm gradually extracts the sparse components of the signals by thresholding the estimated signals iteratively. Each iteration involves two different steps of thresholding and projection. The thresholding step provides an approximation of the common support of the signals by hard-thresholding the summation of the absolute values of the approximated signals. The projection step projects each of the estimated signals onto the convex set defined by the support vector approximated in the previous step.

    In this algorithm $\lambda $  is the relaxation parameter and controls the convergence speed. The threshold value is decreased exponentially by $\beta {e^{ - \alpha \left( {k - 1} \right)}}$, where $k$ is the iteration number. The algorithm performance is less dependent on the parameters $\lambda $, $\beta $,  and $\alpha $; however, these parameters are optimized empirically to achieve a faster convergence.

\begin{algorithm}
\caption{SIMAT}
\label{alg:1}
\begin{algorithmic}[1]
\State \textbf{Input}:
\State \quad Measurement matrices: $\mathbf{A}=[\mathbf{A}^{1}, \mathbf{A}^{2}, ..., \mathbf{A}^{L}]\in \mathbb{R}^{m\times (L\times n)}$
\State \quad Measurement vectors: $\mathbf{Y}=[\mathbf{y}^{1}, \mathbf{y}^{2}, ..., \mathbf{y}^{L}] \in \mathbb{R}^{m\times L}$
\State \quad Maximum number of iterations: $K$
\State \quad three constants: $\alpha,\beta,\lambda$
\State \quad Stopping threshold: $\varepsilon$
\State \textbf{Output}:
\State \quad Recovered estimate of the signals: $\mathbf{\hat{X}} = \mathbf{\hat{x}}^1, \mathbf{\hat{x}}^2, ..., {\mathbf{\hat{x}}}^L \in \mathbb{R}^{n\times L}$
\Procedure{}{}
\State ${\bf{x}}_0^i \leftarrow {\left( {{{\bf{A}}^i}} \right)^\dag } \cdot {{\bf{y}}^i}\;\;1 \le i \le L$, \quad $k\leftarrow1$
\While {$e > \varepsilon \;\& \;k < K$}
\State ${\bf{xsum}} \leftarrow 0$
\State ${\bf{sup}}{{\bf{p}}_k} \leftarrow {\rm{TH}}\left( {{\bf{xsum}},\beta {e^{ - \alpha \left( {k - 1} \right)}}{\rm{\;}}} \right)$
\For{$1\leq i\leq L$}
\State ${\mathbf{s}}_k^i = {\bf{sup}}{{\mathbf{p}}_k} \times {\mathbf{x}}_{k - 1}^i$
\State ${\mathbf{x}}_k^i \leftarrow {\mathbf{s}}_k^i + \;\lambda \left( {{\mathbf{x}}_0^i - {{\left( {{{\mathbf{A}}^i}} \right)}^\dag }{{\mathbf{A}}^i}{\mathbf{s}}_k^i} \right)$.
\State ${\bf{xsum}} \leftarrow {\bf{xsum}} + \frac{{\left| {{\mathbf{x}}_k^i} \right|}}{L}$
\EndFor
\State $k \leftarrow k + 1$
\State $e \leftarrow \|{{\bf{X}}_k} - {{\bf{X}}_{k - 1}}\|$
\EndWhile
\State\Return $\mathbf{\hat{X}} = \mathbf{x}^1_K, \mathbf{x}^2_K, ..., {\mathbf{x}}_K^L$
\EndProcedure
\end{algorithmic}
\end{algorithm}

\subsection{Analytical Discussion}\label{analysis}
    In \cite{esmaeili2016iterative}, it is proved that the IMAT method converges to the sparset solution of the random sample SMV problem if the threshold value is less than
    $\frac{\varepsilon^{2}}{k}$, where k is the sparsity number, and $\varepsilon$ is the minimum distance between the subspace induced by the random samples and all the subspaces induced by a specific support which do not intersect with the previously mentioned subspace. Now, we show that the probability of finding the support of signals with the SIMAT is more than that of the IMAT. Before proving this statement, we note the following points.

    We assume that the original signals are sparse in the time domain, and each non-zero coefficient has a Gaussian distribution. Additionally, the signals are random-sampled and polluted by additive white Gaussian noise in the frequency domain. Since the signals are estimated by line 16 of Algorithm \ref{alg:1} in each iteration, each coefficient of the estimated signals has a Gaussian distribution.

    The distribution of the absolute value of  a random variable with Gaussian distribution $N(0,\sigma^2)$ is Half-Normal with the following Probability Density (PDF) and Cumulative Distribution (CDF):

\begin{equation}\label{eq:4}
{P_Y}\left( {y;\sigma } \right) = \;\frac{{\sqrt 2 }}{{\sqrt {{\sigma ^2} \times \pi } }}\exp \left( { - \frac{{{y^2}}}{{2 \times {\sigma ^2}}}} \right)\;\;\;\;y \ge 0
\end{equation}

\begin{equation}\label{eq:5}
{Q_Y}\left( {y;\sigma } \right) = {\rm{erf}}\left( {\frac{y}{{\sqrt {{\sigma ^2} \times 2} }}} \right)\;\;
\end{equation}

In the thresholding step of each iteration of our algorithm, the absolute values of the coefficients are calculated. Hence, these variables have Half-Normal distribution. We define the variable $z$ as the absolute value of the estimation of a non-zero coefficient of a signal and assume that it has variance ${\sigma ^2}_1$ and mean ${\mu _1}$. Similarly, we define the variable $w$ associated with a zero-coefficient of a signal and assume that its variance and mean are ${\sigma ^2}_0$ and ${\mu _0}$, respectively. We also assume that ${\mu _1} > {\mu _0}$ since the variance of the additive noise is small. Due to the fact that the mean of $L$ signals is calculated for the SIMAT, the variables $z_L$ and $w_L$ can be defined as the mean of the variables mentioned above.  The variance and the mean of $z_L$ and $w_L$ are $\frac{{{\sigma ^2}_1}}{L}$, ${\mu _1}$,
$\frac{{{\sigma ^2}_0}}{L}$ and ${\mu _0}$, as a result of the independency of signals.

    We prove the superiority of SIMAT over IMAT through Lemma \ref{lem:1} and Lemma \ref{lem:2}. As a general idea of these lemmas, according to the law of large numbers, the coefficients of the signal in the SIMAT become nearer to their mean when $L$, the number of signals, go to infinity. In fact, the none-zero coefficients tend to ${\mu _1}$ and the zero coefficients tend to ${\mu _0}$. Therefore, it would be easier to find the non-zero coefficients of the original signals.

\begin{lemma}\label{lem:1}
If the number of signals $L$ satisfies the following inequality
\begin{equation}\label{eq:6}
L \ge \frac{{{\sigma ^2}_0}}{{\left( {\varepsilon  - {\mu _0}} \right)\left( {1 - {\rm{erf}}\left( {\frac{\varepsilon }{{\sqrt {{\sigma ^2}_0 \times 2} }}} \right)} \right)}} > 0,
\end{equation}
then $P\left( {w \ge \varepsilon } \right) \ge P\left( {{w_L} \ge \varepsilon } \right),\;\forall \varepsilon  \ge 2 \times {\mu _0}$, where $P$ is the probability sign. $w$ and $w_L$ are the variables representing the estimated zero coefficients of a signal in each iteration of IMAT and SIMAT, respectively.
\end{lemma}

\begin{IEEEproof}
As a result of $\varepsilon  \ge 2 \times {\mu _0}$ and positivity of $w_L$, one can easily show the following equality:

\begin{equation}\label{eq:7}
P\left( {{w_L} \ge \varepsilon } \right) = \;P\left( {\left| {{w_L} - {\mu _{0\;}}} \right| \ge \varepsilon  - {\mu _0}} \right)
\end{equation}

According to the Chebyshev's theorem, we have

\begin{equation}\label{eq:8}
P\left( {\left| {{w_L} - {\mu _0}} \right| \ge \varepsilon  - {\mu _0}} \right) \le \frac{{\frac{{{\sigma ^2}_0}}{L}}}{{\varepsilon  - {\mu _0}}}
\end{equation}

If we define ${Q_w}\left( \varepsilon  \right)$ as the CDF of $P\left( w \right)$, we get

\begin{equation}\label{eq:9}
P\left( {w \ge \varepsilon } \right) = 1 - {Q_w}\left( \varepsilon  \right) = 1 - {\rm{erf}}\left( {\frac{\varepsilon }{{\sqrt {{\sigma ^2}_0 \times 2} }}} \right)
\end{equation}

Since $1 - {\rm{erf}}\left( {\frac{\varepsilon }{{\sqrt {{\sigma ^2}_0 \times 2} }}} \right) \ge 0$ and $\left( {\varepsilon  - {\mu _0}} \right) \ge 0$, the lemma is proved for the number of signals specified by (\ref{eq:6}).

\end{IEEEproof}

\begin{lemma}\label{lem:2}
If the number of signals $L$ satisfies the following inequality
\begin{equation}\label{eq:10}
L \ge \frac{{{\sigma ^2}_1}}{{\left( {{\mu _1} - \varepsilon } \right)\left( {{\rm{erf}}\left( {\frac{\varepsilon }{{\sqrt {{\sigma ^2}_1 \times 2} }}} \right)} \right)}} > 0,
\end{equation}
then $P\left( {z \ge \varepsilon } \right) \le P\left( {{z_L} \ge \varepsilon } \right),\;\forall \;\;{\mu _1} \ge \varepsilon  \ge 2 \times {\mu _0}$, where $P$ is the probability sign, and the estimated non-zero coefficients of a signal in each iteration of IMAT and SIMAT are denoted by $z$ and $z_L$, respectively.
\end{lemma}

\begin{IEEEproof}
It is not too difficult to derive the following inequality

\begin{equation}\label{eq:11}
P\left( {{z_L} \ge \varepsilon } \right) \ge 1 - \;P\left( {\left| {{z_L} - {\mu _1}} \right| \ge {\mu _1} - \varepsilon } \right)
\end{equation}

According to the Chebyshev's theorem, we have

\begin{equation}\label{eq:12}
\begin{split}
  & P\left( {\left| {{z_L} - {\mu _1}} \right| \ge {\mu _1} - \varepsilon } \right) \le \frac{{\frac{{{\sigma ^2}_1}}{L}}}{{{\mu _1} - \varepsilon }}
 \Rightarrow \\
     & P\left( {{z_L} \ge \varepsilon } \right) \ge 1 - \;\frac{{\frac{{{\sigma ^2}_1}}{L}}}{{{\mu _1} - \varepsilon }}
\end{split}
\end{equation}

If we define ${Q_z}\left( \varepsilon  \right)$ as the CDF of $P(z)$, then we get

\begin{equation}\label{eq:13}
P\left( {z \ge \varepsilon } \right) = 1 - {Q_z}\left( \varepsilon  \right) = 1 - {\rm{erf}}\left( {\frac{\varepsilon }{{\sqrt {{\sigma ^2}_1 \times 2} }}} \right)
\end{equation}

Since ${\rm{erf}}\left( {\frac{\varepsilon }{{\sqrt {{\sigma ^2}_1 \times 2} }}} \right) \ge 0$ and $\left( {{\mu _1} - \varepsilon } \right) \ge 0$ for ${\mu _1} \ge \varepsilon  \ge 2 \times {\mu _0}$, one can easily prove the lemma for the $L$ indicated in (\ref{eq:10})

\end{IEEEproof}

\begin{theorem}\label{th:1}
If the number of signals $L$ satisfies the following inequality
\begin{equation}\label{eq:14}
\begin{split}
   \forall \;\;\;{\mu _1} &\ge \varepsilon\ge 2 \times {\mu _0}\;\;\;\;, \\
     & L \ge {\rm{max}}(\frac{{{\sigma ^2}_1}}{{\left( {{\mu _1} - \varepsilon } \right)\left( {{\rm{erf}}\left( {\frac{\varepsilon }{{\sqrt {{\sigma ^2}_1 \times 2} }}} \right)} \right)}} \\
     & ,\frac{{{\sigma ^2}_0}}{{\left( {\varepsilon  - {\mu _0}} \right)\left( {1 - {\rm{erf}}\left( {\frac{\varepsilon }{{\sqrt {{\sigma ^2}_0 \times 2} }}} \right)} \right)}}),
\end{split}
\end{equation}
then $\forall \;\;{\mu _1} \ge \varepsilon  \ge 2 \times {\mu _0}$, the probability of finding the support with SIMAT is higher than IMAT.
\end{theorem}

\begin{IEEEproof}
According to Lemma \ref{lem:1}, if L satisfies (\ref{eq:6}), the probability of mistaking a zero coefficient for a support in SIMAT is less than that of IMAT.
According to Lemma \ref{lem:2}, if L satisfies (\ref{eq:10}), the probability of finding a none-zero coefficients in SIMAT is higher than IMAT.

Therefore, if $L$ satisfies (\ref{eq:14}), it is more probable to find the support with the SIMAT than with the IMAT.
\end{IEEEproof}

One can find the minimum value of $L$ which satisfies (\ref{eq:14}) as shown in Table \ref{tab:4}. This table shows that the minimum number of signals $L$ that guarantees superiority of SIMAT over IMAT is very low, and hence there is no need to have a large number of signals to benefit from SIMAT.

\begin{table}
\renewcommand{\arraystretch}{1.3}
\caption[caption]{Minimum Number of Signals that Satisfies Theorem \ref{th:1}.}
\label{tab:4}
\centering
\begin{tabular}{|c||c|c|c|}
    \hline

    \textbf{Input SNR (dB)} & \textbf{10} & \textbf{20} & \textbf{50} \\
    \hline
    \hline

    \textbf{$\mu_{0}$} & 0.25 & 0.079 & 0.0025\\
    \hline

    \textbf{$\mu_{1}$} & 0.5 & 0.5 & 0.5\\
    \hline

    \textbf{${\sigma_{0}}^{2}$} & 0.1 & 0.01 & $10^{-5}$\\
    \hline

    \textbf{${\sigma_{1}}^{2}$} & 0.18 & 0.09 & 0.08\\
    \hline

    \textbf{$\varepsilon$} & 0.5 & 0.2 & 0.011\\
    \hline

    \textbf{Minimum Value of $L$} & \textbf{4} & \textbf{2} & \textbf{5}\\
    \hline

\end{tabular}
\end{table}

\section{Simulation Results}\label{simulation}

\begin{figure}
  \centering
  \includegraphics[width=3.2in]{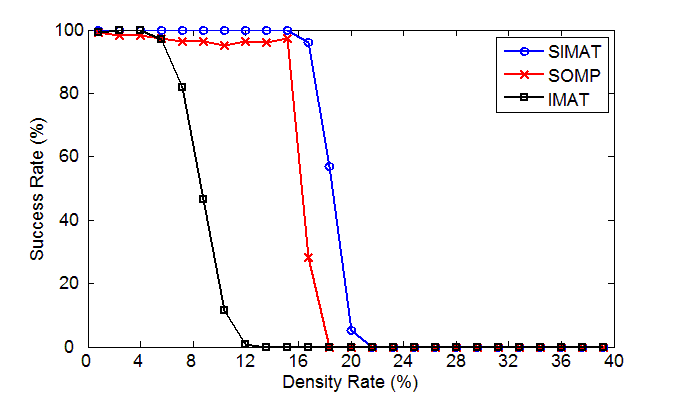}
  \caption{Success Rate vs. Density Rate (Input SNR$=20$dB,  Sampling Rate = $25\%$,  $L=8$, and the Sparsity Number is Unknown).}\label{fig:1}
\end{figure}

\begin{figure}
\centering
\includegraphics[width=3.2in]{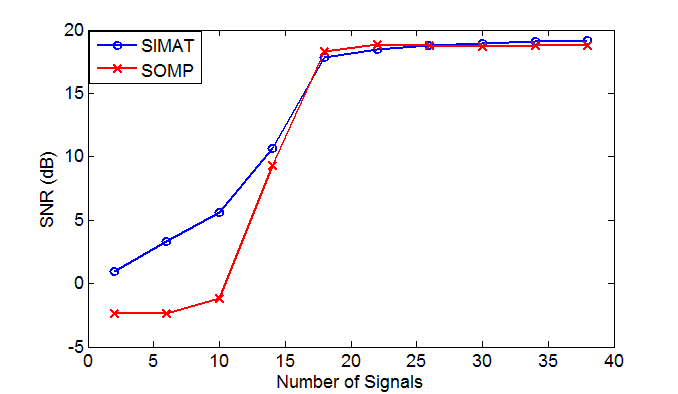}
\caption{Reconstruction SNR vs. Number of Signals (Input SNR=$20$dB,  Sampling Rate = $25\%$, Density Rate $K=20\%$, and the Sparsity Number is Unknown).}\label{fig:2}
\end{figure}

\begin{table*}[htbp]
\renewcommand{\arraystretch}{1.3}
\caption[caption]{Reconstruction SNR in $\mathrm{dB}$ of Three Different Algorithms for Different Density Rate (K (\%))
\\in the Presence of Noise (SNR0 = Input SNR)(Sampling Rate=25\%, $L$=8)}
\label{tab:1}
\centering
\begin{tabular}{c|c|c|c|c|c|c|c|c|c|c|c|c|c|c|c|c|c|c|}
    \cline{2-19}
     & \multicolumn{9}{ c|| }{ \textbf{K is Unknown}} & \multicolumn{9}{ c| }{ \textbf{K is Known}} \\
    \hline
    \hline

    \multicolumn{1}{ |c|| }{ \textbf{SNR}\textsubscript{\textbf{0}} \textbf{(dB)}} & \multicolumn{3}{ c| }{ \textbf{10}} & \multicolumn{3}{ c| }{ \textbf{20}} & \multicolumn{3}{ c|| }{ \textbf{100}} & \multicolumn{3}{ c| }{ \textbf{10}} & \multicolumn{3}{ c| }{ \textbf{20}} & \multicolumn{3}{ c| }{ \textbf{100}} \\
    \hline

    \multicolumn{1}{ |c|| }{ \textbf{K (\%)}} & \textbf{4} & \textbf{12} & \textbf{20} & \textbf{4} & \textbf{12} & \textbf{20} & \textbf{4} & \textbf{12} & \multicolumn{1}{ |c|| }{ \textbf{20}} & \textbf{4} & \textbf{12} & \textbf{20} & \textbf{4} & \textbf{12} & \textbf{20} & \textbf{4} & \textbf{12} & \textbf{20}\\
    \hline
    \hline


    \multicolumn{1}{ |c|| }{\textbf{IMAT}} & 19.7 & 0.5 & -1.3 & 31.4 & 0.7 & -1.8 & 94.3 & 0.9 & \multicolumn{1}{ |c|| }{ -2.1} & 19.9 & 2.1 & 1.3 & 31.4 & 2.2 & -0.5 & 94.6 & 1.4 & -1.9\\
    \hline

    \multicolumn{1}{ |c|| }{\textbf{SIMAT}} & 23.2 & 16.1 & 1.8 & 31.1 & 25.4 & 4.6 & 106 & 105.8 & \multicolumn{1}{ |c|| }{ 14.5} & 23.3 & 17.2 & 3.7 & 31.3 & 26.4 & 5.9 & 111.3 & 106.8 & 16.6\\
    \hline

    \multicolumn{1}{ |c|| }{\textbf{SOMP}} & 14.9 & 12.9 & 1.2 & 23.1 & 23.1 & 2.1 & 114.1 & 107.3 & \multicolumn{1}{ |c|| }{ 22.5} & 23 & 16.8 & 0.4 & 32.9 & 26.8 & 4.7 & 113.9 & 107.4 & 21.4\\
    \hline

\end{tabular}
\end{table*}

\begin{table}
\renewcommand{\arraystretch}{1.3}
\caption[caption]{Run-time in Seconds of Different Algorithms for Different Density Rate (K (\%)) in the Presence of Noise \\(Input SNR = 100 $\mathrm{dB}$), (Sampling Rate=25\%, $L$=8 , K is known).}
\label{tab:2}
\centering
\begin{tabular}{|c||c|c|c|}
    \hline

    \textbf{K (\%)} & \textbf{4} & \textbf{12} & \textbf{20} \\
    \hline
    \hline


    \textbf{SIMAT} & 0.017 & 0.027 & 0.027\\
    \hline

    \textbf{SOMP} & 0.031 & 0.081 & 0.138\\
    \hline

\end{tabular}
\end{table}

\begin{figure}
\centering
\includegraphics[width=3.2in]{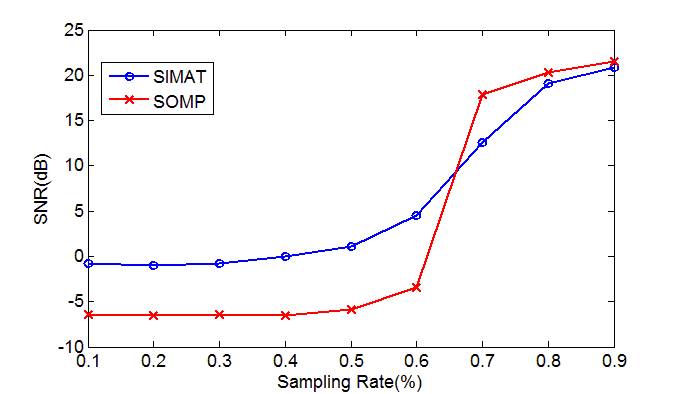}
\caption{Reconstruction SNR vs. Sampling Rate (Input SNR=$20$dB, Density Rate $K=60\%, L=8$,and the Sparsity Number is Unknown).}\label{fig:3}
\end{figure}

In each trial of our simulation, we generate $L$ number of $K$-sparse signals. We choose $K$ components out of$ N=256$ randomly, and set them to a random number in the interval $[-1,1]$. Then the noisy signals are random-sampled by a sampling rate of $M/N$. By the law of algebra, the number of samples needed to specify the sparsity profile of the signals is at least twice the sparsity number, hence $K \leq M/((2\times N))$. We optimize the parameters of the algorithm in each trial.

Table \ref{tab:1} compares the average reconstruction SNR (dB) of three algorithms, IMAT, SIMAT and SOMP, for different density numbers and input SNRs. As observed in Table \ref{tab:1}, the simultaneous reconstruction methods outperform IMAT, especially for higher sparsity numbers.

In the case of not knowing the density rate, SIMAT yields the best results both in low and high density rate, and in noisier channels. In noiseless channels, the signal can be perfectly reconstructed by SIMAT, when the sparsity number is small.

In the case of knowing the density rate, SOMP and SIMAT exhibit similar performances in all the cases. However, based on the results of Table \ref{tab:2}, SIMAT outperforms SOMP in terms of the complexity measured by the run-time.

 The success rate of IMAT, SIMAT and SOMP algorithms for different density rates is depicted in Fig. \ref{fig:1}. A reconstruction is considered to be successful if the output SNR is more than 20dB. As seen in this figure, all curves experience a sudden knee-like fall as the density rate increases. This fall is considered as the boundary between successful and unsuccessful reconstruction. The simulation results reveal that for the SIMAT, success rate falls around $20\%$ density rate. While, the knee-like fall happens in $18\%$ and $12\%$ density rate for IMAT and SOMP, respectively. This indicates that SIMAT can successfully reconstruct the signals with higher sparsity number in comparison with SOMP and IMAT algorithms.

Fig. \ref{fig:2} shows the effect of the number of signals on the performance of SOMP and SIMAT.  As observed in this figure, by increasing the number of signals, we get better results. Additionally,
the SIMAT algorithm can reconstruct the signals better than the SOMP when the number of signals is small.

  The reconstruction SNR for different sampling rates is depicted in Fig. \ref{fig:3}. According to this figure, the SNR values can be improved by increasing the sampling rate. Although the SOMP algorithm has better results for a small interval in the high sampling rates, SIMAT outperforms SOMP in most cases.

   \begin{table}
\renewcommand{\arraystretch}{1.3}
\caption[caption]{Reconstruction SNR in $\mathrm{dB}$ of Different Algorithms for \\Different Density Rate (K (\%))
in the Presence of Nois \\(Input SNR=5 $\mathrm{dB}$, Sampling rate=20\% , $L$=10).}
\label{tab:3}
\centering
\begin{tabular}{|c||c|c|c|c|c|c|}
    \hline

    \textbf{K (\%)} & \textbf{2.5} & \textbf{5} & \textbf{7.5} & \textbf{10} & \textbf{12.5} & \textbf{15}\\
    \hline
    \hline


    \textbf{SIMAT} & 14.1 & 11.6 & 8.8 & 6.7 & 4.9 & 3.3\\
    \hline

    \textbf{BIMAT} & 13.3 & 9.8 & 7.5 & 6.0 & 4.1 & 3.4\\
    \hline

    \textbf{Improvement Ratio(\%)}& +6 & +18 & +17 & +12 & +19 & -2\\
    \hline

\end{tabular}
\end{table}

   Table \ref{tab:3} lists the SNR of two reconstruction methods extended from IMAT. Simulation results demonstrate the superiority of SIMAT over BIMAT in terms of the output SNR and the complexity. Therefore, the SIMAT can be used instead of the BIMAT in many applications such as distributed MIMO radar systems.

\section{Conclusion}\label{conclusion}
In this paper a novel method, SIMAT, was introduced for SSA problems. The proposed method is an extension of IMAT into the MMV models, and the idea of this extension is that the summation of sparse vectors, sharing the same support, in the thesholding step of each iteration can enhance the probability of reconstruction. Indeed, the theoretical analysis with simulation results prove that SIMAT outperforms IMAT with respect to the SNR metric. We compared the proposed method with SOMP, as a well-known algorithm in the MMV models. We can conclude from the conducted numerical experiments that SIMAT is preferable in terms of SNR or SR, specially in noisier channels with low sampling rates and high density rates. Finally, it was observed that SIMAT is superior to BIMAT when complexity and efficiency are important factors.

\section*{Acknowledgment}
The authors would like to thank Ehsan Asadi for helping in the analytical discussion subsection and Babak Barazandeh for introducing this topic to one of the authors.


\begin{thebibliography}{10}
\providecommand{\url}[1]{#1}
\csname url@samestyle\endcsname
\providecommand{\newblock}{\relax}
\providecommand{\bibinfo}[2]{#2}
\providecommand{\BIBentrySTDinterwordspacing}{\spaceskip=0pt\relax}
\providecommand{\BIBentryALTinterwordstretchfactor}{4}
\providecommand{\BIBentryALTinterwordspacing}{\spaceskip=\fontdimen2\font plus
\BIBentryALTinterwordstretchfactor\fontdimen3\font minus
  \fontdimen4\font\relax}
\providecommand{\BIBforeignlanguage}[2]{{%
\expandafter\ifx\csname l@#1\endcsname\relax
\typeout{** WARNING: IEEEtran.bst: No hyphenation pattern has been}%
\typeout{** loaded for the language `#1'. Using the pattern for}%
\typeout{** the default language instead.}%
\else
\language=\csname l@#1\endcsname
\fi
#2}}
\providecommand{\BIBdecl}{\relax}
\BIBdecl

\bibitem{candes2006robust}
E.~J. Cand{\`e}s, J.~Romberg, and T.~Tao, ``Robust uncertainty principles:
  Exact signal reconstruction from highly incomplete frequency information,''
  \emph{IEEE Transactions on information theory}, vol.~52, no.~2, pp. 489--509,
  2006.

\bibitem{donoho2006compressed}
D.~L. Donoho, ``Compressed sensing,'' \emph{IEEE Transactions on information
  theory}, vol.~52, no.~4, pp. 1289--1306, 2006.

\bibitem{duarte2009model}
M.~F. Duarte, V.~Cevher, and R.~G. Baraniuk, ``Model-based compressive sensing
  for signal ensembles,'' in \emph{Communication, Control, and Computing, 2009.
  Allerton 2009. 47th Annual Allerton Conference on}.\hskip 1em plus 0.5em
  minus 0.4em\relax IEEE, 2009, pp. 244--250.

\bibitem{gastpar2003source}
M.~Gastpar and M.~Vetterli, ``Source-channel communication in sensor
  networks,'' in \emph{Information Processing in Sensor Networks}.\hskip 1em
  plus 0.5em minus 0.4em\relax Springer, 2003, pp. 553--553.

\bibitem{gorodnitsky1995neuromagnetic}
I.~F. Gorodnitsky, J.~S. George, and B.~D. Rao, ``Neuromagnetic source imaging
  with focuss: a recursive weighted minimum norm algorithm,''
  \emph{Electroencephalography and clinical Neurophysiology}, vol.~95, no.~4,
  pp. 231--251, 1995.

\bibitem{malioutov2005sparse}
D.~Malioutov, M.~Cetin, and A.~S. Willsky, ``A sparse signal reconstruction
  perspective for source localization with sensor arrays,'' \emph{IEEE
  transactions on signal processing}, vol.~53, no.~8, pp. 3010--3022, 2005.

\bibitem{ibernon2008comparison}
R.~Ibernon-Fernandez, J.-M. Molina-Garcia-Pardo, and L.~Juan-Llacer,
  ``Comparison between measurements and simulations of conventional and
  distributed mimo system,'' \emph{IEEE Antennas and Wireless Propagation
  Letters}, vol.~7, pp. 546--549, 2008.

\bibitem{mohammadi2017sampling}
E.~Mohammadi, A.~Fallah, and F.~Marvasti, ``Sampling and distortion tradeoffs
  for indirect source retrieval.''\hskip 1em plus 0.5em minus 0.4em\relax IEEE,
  2017, accepted for publication.

\bibitem{tropp2007signal}
J.~A. Tropp and A.~C. Gilbert, ``Signal recovery from random measurements via
  orthogonal matching pursuit,'' \emph{IEEE Transactions on information
  theory}, vol.~53, no.~12, pp. 4655--4666, 2007.

\bibitem{determe2017noise}
J.-F. Determe, J.~Louveaux, L.~Jacques, and F.~Horlin, ``On the noise
  robustness of simultaneous orthogonal matching pursuit,'' \emph{IEEE
  transactions on signal processing}, vol.~65, no.~4, pp. 864--875, 2017.

\bibitem{tropp2006algorithms}
J.~A. Tropp, A.~C. Gilbert, and M.~J. Strauss, ``Algorithms for simultaneous
  sparse approximation. part i: Greedy pursuit,'' \emph{Signal Processing},
  vol.~86, no.~3, pp. 572--588, 2006.

\bibitem{rakotomamonjy2011surveying}
A.~Rakotomamonjy, ``Surveying and comparing simultaneous sparse approximation
  (or group-lasso) algorithms,'' \emph{Signal processing}, vol.~91, no.~7, pp.
  1505--1526, 2011.

\bibitem{marvasti2012sparse}
F.~Marvasti, M.~Azghani, P.~Imani, P.~Pakrouh, S.~J. Heydari, A.~Golmohammadi,
  A.~Kazerouni, and M.~Khalili, ``Sparse signal processing using iterative
  method with adaptive thresholding (imat),'' in \emph{Telecommunications
  (ICT), 2012 19th International Conference on}.\hskip 1em plus 0.5em minus
  0.4em\relax IEEE, 2012, pp. 1--6.

\bibitem{marvasti2012unified}
F.~Marvasti, A.~Amini, F.~Haddadi, M.~Soltanolkotabi, B.~H. Khalaj,
  A.~Aldroubi, S.~Sanei, and J.~Chambers, ``A unified approach to sparse signal
  processing,'' \emph{EURASIP journal on advances in signal processing}, vol.
  2012, no.~1, p.~44, 2012.

\bibitem{marvasti2012nonuniform}
F.~Marvasti, \emph{Nonuniform sampling: theory and practice}.\hskip 1em plus
  0.5em minus 0.4em\relax Springer Science \& Business Media, 2012.

\bibitem{abtahi2016iterative}
A.~Abtahi, M.~Azghani, J.~Tayefi, and F.~Marvasti, ``Iterative block-sparse
  recovery method for distributed mimo radar,'' in \emph{Communication and
  Information Theory (IWCIT), 2016 Iran Workshop on}.\hskip 1em plus 0.5em
  minus 0.4em\relax IEEE, 2016, pp. 1--4.

\bibitem{esmaeili2016iterative}
A.~Esmaeili, E.~Asadi, and F.~Marvasti, ``Iterative null-space projection
  method with adaptive thresholding in sparse signal recovery and matrix
  completion,'' \emph{arXiv preprint arXiv:1610.00287}, 2016.

\end{thebibliography}
\end{document}